%% file: Ricalib_main.tex
\documentclass[journal]{IEEEtran}
\usepackage{graphicx}
\usepackage[thickspace,amssymb]{siunits}
\begin{document}
\title{Extended calibration range for  prompt photon emission in ion beam irradiation.  }
\author{
F.~Bellini$^{a,b}$, 
T.T.~Boehlen$^{c}$, 
M.P.W.~Chin$^{c}$,
F.~Collamati$^{a,b}$, 
E.~De~Lucia$^{d}$, 
R.~Faccini$^{a,b}$, 
A.~Ferrari$^{c}$,
L.~Lanza$^{a,b}$, 
C. Mancini-Terraciano$^{c,f}$,
M.~Marafini$^{e,b}$, 
I.~Mattei$^{f,d}$, 
S.~Morganti$^{b}$, 
P.G.~Ortega$^{c}$,
V.~Patera$^{g,b}$, 
L.~Piersanti$^{g,d}$, 
A.~Russomando$^{h,b}$, 
 P.R.~Sala$^{i}$, 
A.~Sarti$^{g,d}$, 
A.~Sciubba$^{g,b}$, 
E.~Solfaroli~Camillocci$^{h}$, 
C.~Voena$^{b}$ \\ \vspace{0.3cm}
$^a$ Dipartimento di Fisica, Sapienza Universit\`a di Roma, Roma, Italy; 
$^b$ INFN Sezione di Roma, Roma, Italy; 
$^c$ CERN, Geneva, Switzerland;
$^d$ Laboratori Nazionali di Frascati dell'INFN, Frascati, Italy; 
$^e$ Museo Storico della Fisica e Centro Studi e Ricerche ``E.~Fermi'', Roma, Italy; 
$^f$ Dipartimento di Fisica, Universit\`a Roma Tre, Roma, Italy;
$^g$ Dipartimento di Scienze di Base e Applicate per Ingegneria, Sapienza Universit\`a di Roma,  Roma, Italy;
$^h$ Center for Life Nano Science@Sapienza, Istituto Italiano di Tecnologia, Roma, Italy;
$^i$ INFN Sezione di Milano, Milano, Italy.
}
\maketitle
\thispagestyle{empty}

\begin{abstract}
Monitoring the dose delivered during proton and carbon ion therapy is still a matter of research. Among the possible solutions, several exploit the measurement of the  single photon emission from nuclear decays induced by the irradiation. To fully characterize such emission the detectors need development, since the energy spectrum spans the range above the MeV that is not traditionally used in medical applications. On the other hand, a deeper
understanding of the reactions involving gamma production is needed in
order to improve the physic models of Monte Carlo codes, relevant for an
accurate prediction of the prompt-gamma energy spectrum.This paper  describes a calibration technique tailored for the range of energy of interest and reanalyzes the data of the interaction of a  $80\ \mega\electronvolt/$u fully stripped carbon ion beam with a  Poly-methyl methacrylate target. By adopting the FLUKA simulation with the appropriate calibration and resolution a significant improvement in the  agreement between data and simulation is reported.
\end{abstract}

%%%%%%%%%%%%%%%%%%%%%%%%%%%%%%%%%%%%%%%%%%%%%%%%%%%%%%%%%%%%%%%%%%%%%%%%%%%%%%

\input{Ricalib_introduction.tex}

\input{Ricalib_newcalib.tex}

\input{Ricalib_simulation.tex}

\input{Ricalib_results.tex}

\begin{center}\textbf{\large Acknowledgements}\end{center}{\large \par}

The authors would like to thank Dr. M.~Pillon and Dr. M.~Angelone (ENEA-Frascati, Italy) for allowing us to use the ENEA AmBe source.
%%%%%%%%%%%%%%%%%%%%%%%%%%%%%%%%%%%%%%%%%%%%%%%%%%%%%%%%%%%%%%%%%%%%%%%%%%%%%%

%
%

%%%%%%%%%%%%%%%%%%%%%%%%%%%%%%%%%%%%%%%%%%%%%%%%%%%%%%%%%%%%%%%%%%%%%%%%%%%%%%
\end{document}

%% file: Ricalib_introduction.tex
In the last decade, the use of proton and carbon beams has become more and more widespread as an effective therapy for the treatment of solid cancer (hadrontherapy). Due to their very favorable profile of the release dose in tissue, the hadron beams can be very effective in destroying the tumor and sparing the adjacent healthy tissue in comparison to the standard X-ray based treatment~\cite{Amaldi}. On the other hand, the space selectivity of the hadrontherapy asks for a new approach to the delivered dose monitoring.
Indeed, a precise monitoring of the dose is then essential for a good quality control of the treatment. Furthermore, the dose monitoring  would be particularly useful if provided during the treatment (in-beam monitoring) in order to provide a fast quality check of a treatment. %doporef

\begin{figure}[!ht]
\begin{center}
\includegraphics [width = 0.45 \textwidth] {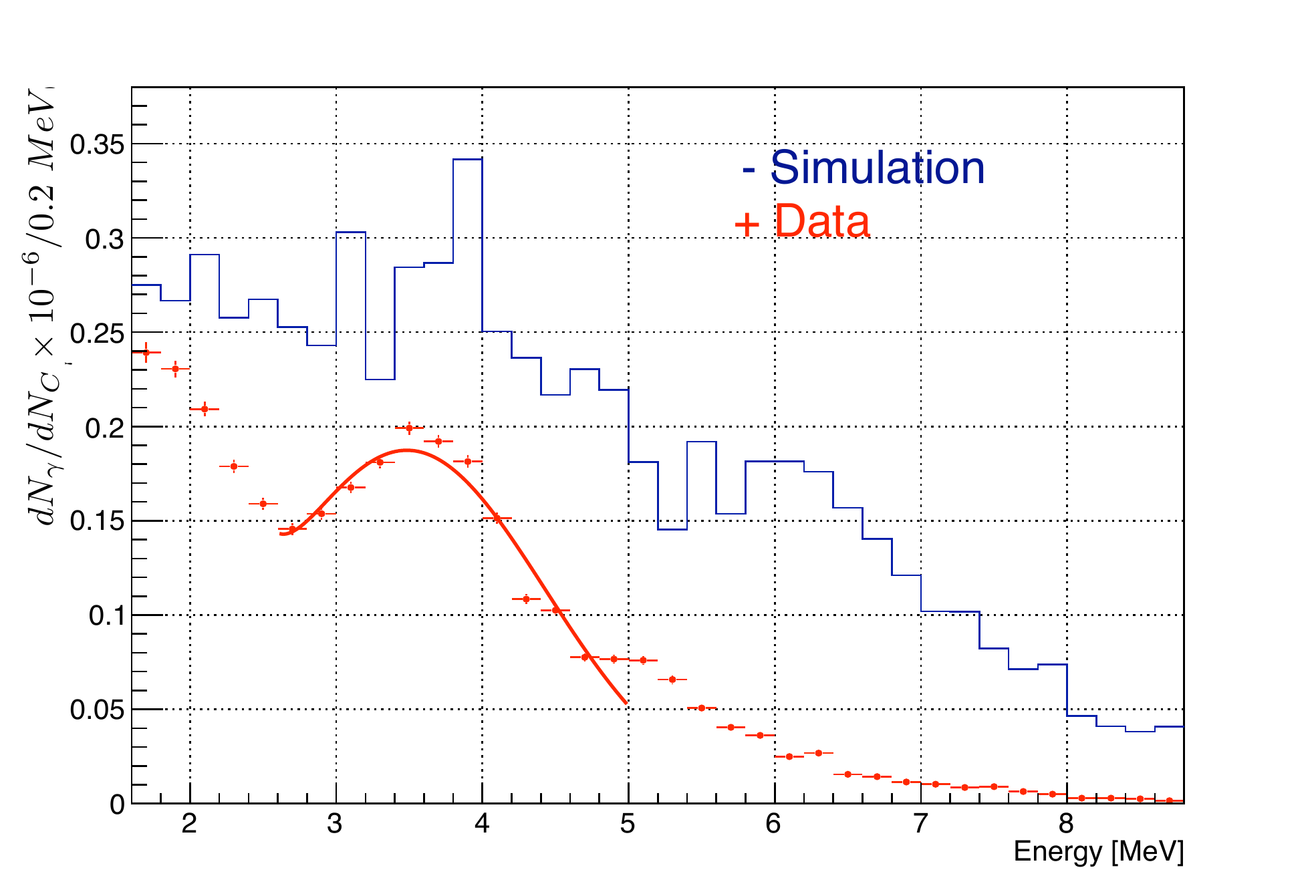}
\caption{Energy spectrum of the prompt photons produced  by irradiating a PMMA target with $80\ \mega\electronvolt/$u fully stripped carbon ions as measured in Ref.~\cite{catania}.  The simulation was performed with the GEANT program~\cite{GEANT}.}
\label{fig:catania}
\end{center}
\end{figure}

Several methods have been developed to determine the Bragg Peak position online by exploiting the secondary particle production induced by the hadron beam. In this paper we concentrate on the ``prompt-photon'' method~\cite{Testa, catania}: since the irradiation of tissues with hadron beams produces nuclear excitations followed by photon emissions from de-excitation within few $\nano\second$ (prompt photons), both rate and production region could be used to monitor the dose release.

The energy spectrum of the emitted radiation was measured by irradiating a Poly-methyl methacrylate (PMMA) target with $80\ \mega\electronvolt/$u fully stripped carbon ions at the Laboratori Nazionali del Sud (LNS) of the Istituto Nazionale di Fisica Nucleare (INFN) in Catania~\cite{catania} and is reported in Fig.~\ref{fig:catania}. Such result has shown that the photons of interest are in the 2-10 MeV range and that the Monte Carlo (MC) program used (GEANT~\cite{GEANT}) does not properly reproduce the main features of the data. Finally, the spectrum also showed that the only visible structure, due to the 4.44 MeV de-excitation line of $^{12}\rm{C}^*$, $^{11}\rm{C}^*$, and $^{11}\rm{B}^*$, is offset in energy with respect  to the expected position.

Photons in this energy range are measured with scintillating crystals read with photo-detectors. In the specific case of Ref.~\cite{catania}, the crystals where chosen to be made of cerium-doped lutetium-yttrium ortho-silicate (LYSO)  in order to ensure the excellent time resolution needed to  discriminate against the relatively slow neutrons. Such crystals were calibrated with the standard $^{22}{\rm Na}$ and $^{60}{\rm Co}$ sources, emitting $511$ keV and $1.17$ plus $1.33$ MeV photons respectively. The calibration in the range of interest was therefore extrapolated at energies higher than the sources could provide.

This paper shows the results of applying  an alternative calibration method based on the detection of monochromatic photons of energies up to 9~MeV  and a comparison between the measured spectra and rates with the FLUKA simulation~\cite{fluka}.

%% file: Ricalib_newcalib.tex
%%%%%%%%%%%%%%%%%%%%%%%%%%%%%%%%%%%%%%%%%%%%%%%%%%%%%%%%%%%%%%%%%%%%%%%%%%%%%%
\section{The extended-range calibration}
\label{newcalib}
A more suited calibration must exploit a source which produces photons of energy above 1 MeV. Given the low lifetime of the isotopes with such lines, an indirect production mechanism needs to be  implemented. We have exploited the possibility to have neutron induced lines and we have used an  AmBe source that produces approximately $2.5\times 10^6$~n/s. The source was hosted inside a 5~cm thick container made of paraffin (C$_{31}$H$_{62}$): this allowed both to moderate the neutron flux, that would otherwise blind the detectors, and to produce $\gamma$ lines. The interaction of the neutrons with the hydrogen produces the 2.22MeV line from deuteron formation (H(n,$\gamma$)d), while the interaction with the carbon produces the 4.44 MeV $^{12}$C$^*$ de-excitation line.

%\begin{figure*}[!ht]
%\begin{center}
%\includegraphics [width = 0.45 \textwidth] {figs/LYSO1-spectrum.pdf}
%\includegraphics [width = 0.45 \textwidth] {figs/LYSO1-calib.pdf}
%\caption{Left: example of energy spectrum as obtained with an AmBe source moderated with paraffin. Right: the corresponding calibration curves assuming a linear (red) or logarithmic (green) behavior. The details of the latter are explained in the text. }
%\label{fig:LYSO1}
%\end{center}
%\end{figure*}

\begin{figure}[!ht]
\begin{center}
\includegraphics [width = 0.45 \textwidth] {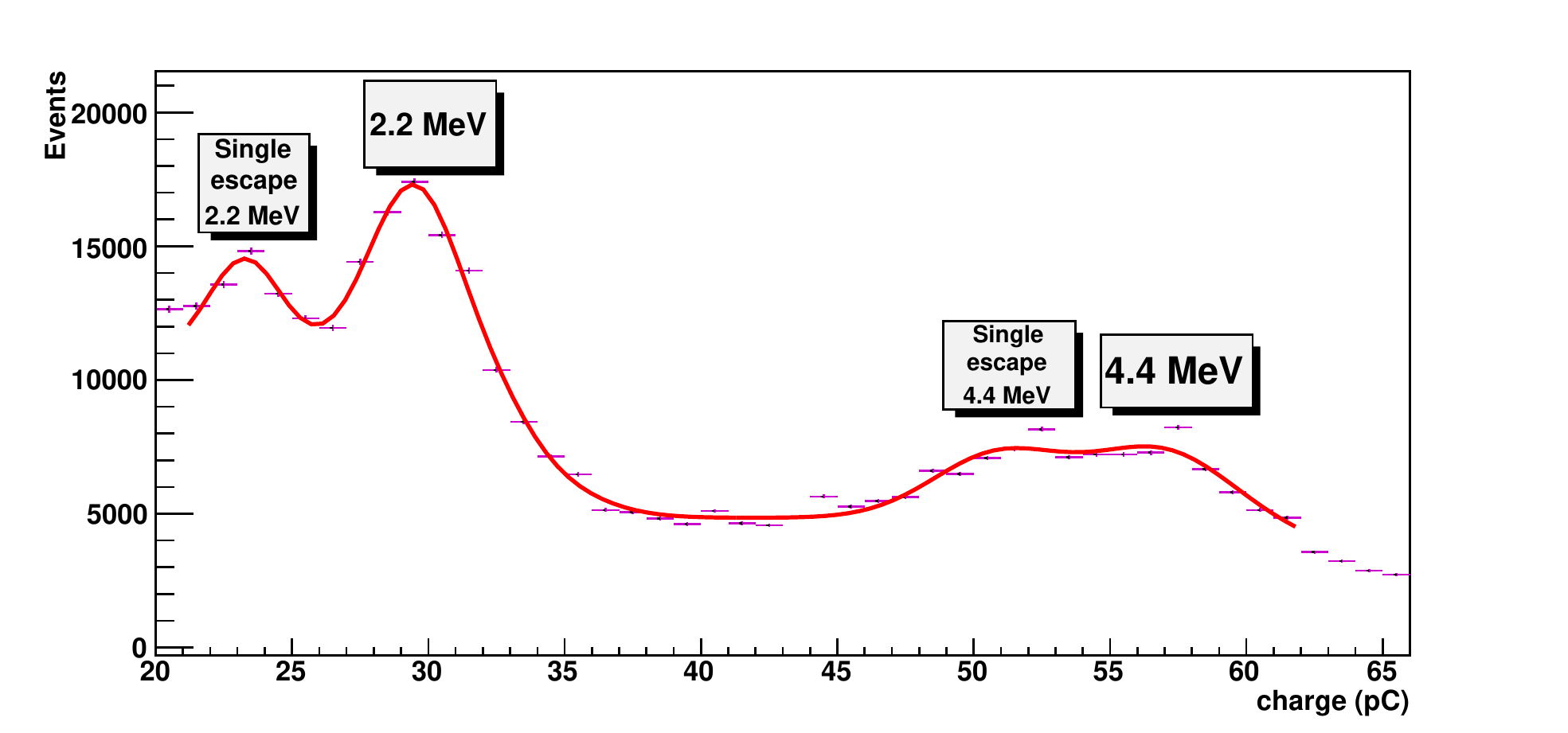}
\includegraphics [width = 0.45 \textwidth] {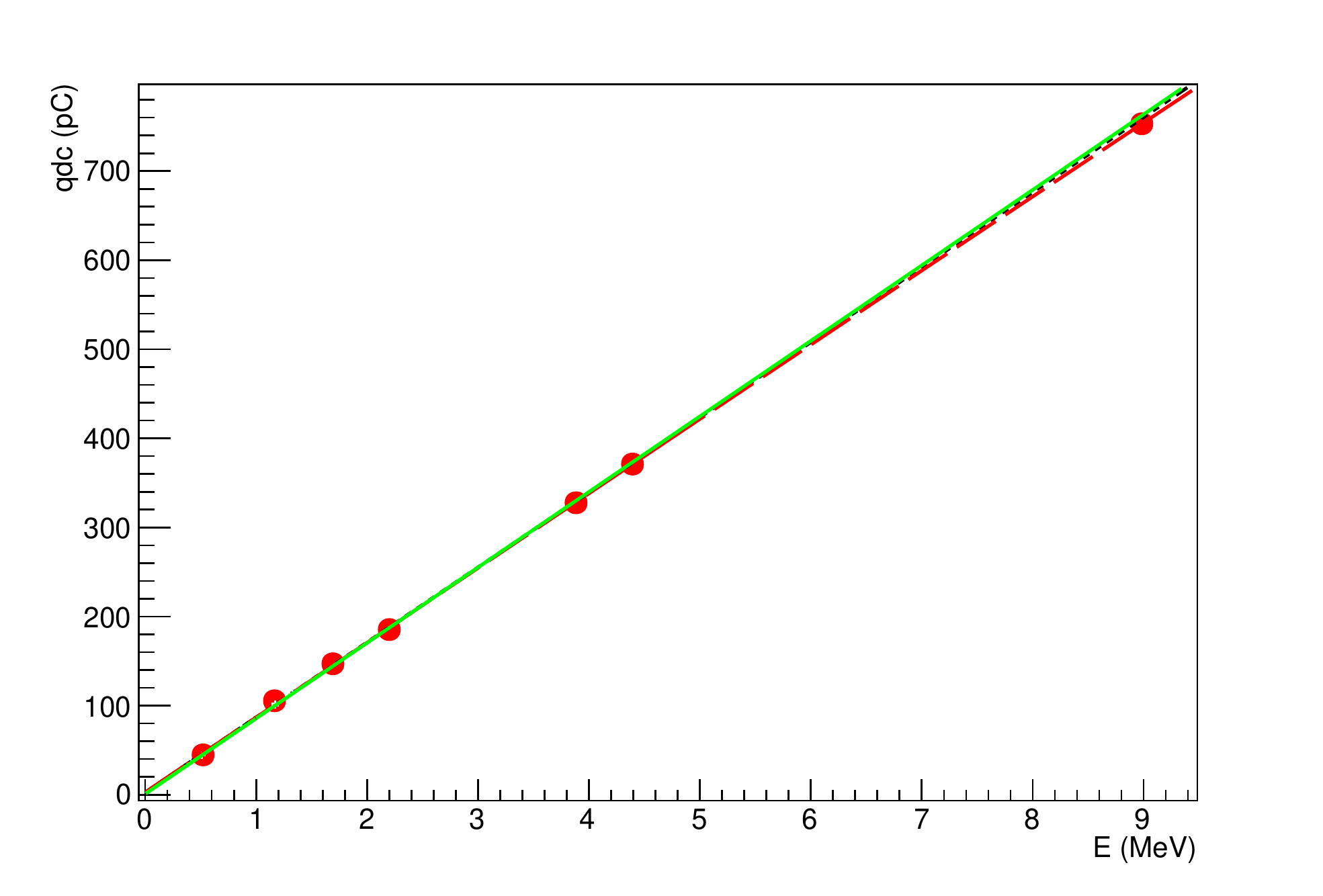}
\caption{Top: example of energy spectrum as obtained with an AmBe source moderated with paraffin with the detector used in Ref.~\cite{catania}. Bottom: the new calibration curve assuming a linear  behavior (dashed black line) compared with the one published in Ref.~\cite{catania} (full green line).  }
\label{fig:LYSO2}
\end{center}
\end{figure}

%This procedure has been first tested on a  detector made of LYSO and a PMT with the same component but lower noise than the one adopted in Ref.~\cite{catania}. 
The measured spectrum is shown in Fig.~\ref{fig:LYSO2}. The 2.22 and 4.44 MeV lines are clearly visible together with  the lines that occur when a 511 keV photon produced by the annihilation of a positron escapes the detector (single escape).  Each possible signal line has been parametrized in the fit as a Gaussian, while the background is a superposition of Fermi-Dirac functions. This allows to assign a measured value of $ADC$ counts to each line and to produce the calibration plot shown in Fig.~\ref{fig:LYSO2} at the bottom. 

\begin{figure}[!ht]
\begin{center}
\includegraphics [width = 0.45 \textwidth] {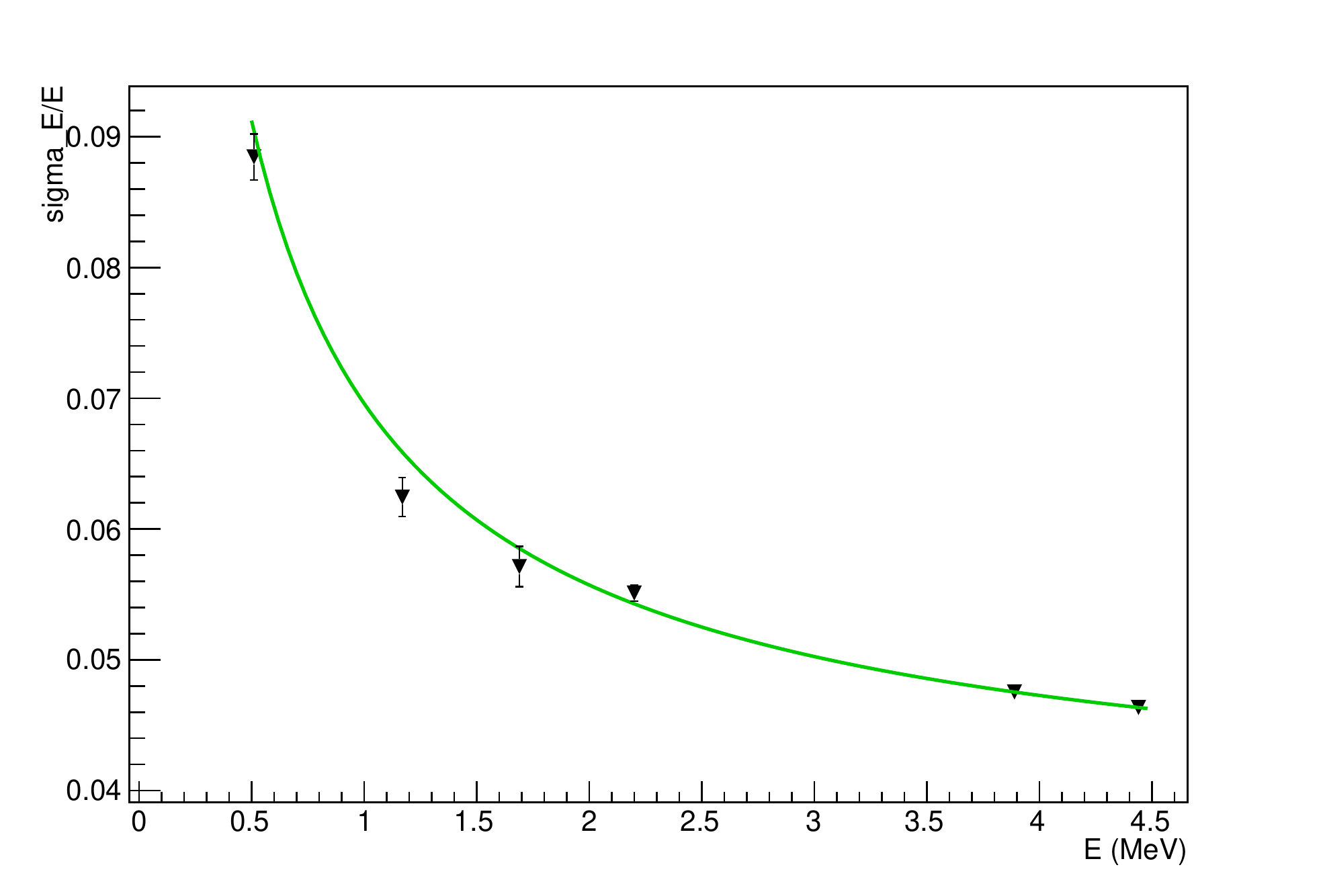}
\caption{ Dependence of the relative fitted resolution on the energy.}
\label{fig:sigma}
\end{center}
\end{figure}

From the fit to the spectrum also the resolution as a function of the energy can be obtained. This information is needed as input to the simulation that will be described in the next paragraph. The result is shown in Fig.~\ref{fig:sigma}, where the dependence is fitted with
\begin{equation}
\frac{\sigma(E)}{E}=\sqrt{p_0^2+\frac{p_1^2}{E}+\frac{p_2^2}{E^2}}.
\end{equation}
We have found the electronic term, $p_2$, to be consistent with zero, the constant term to be $p_0=(3.70\pm0.007)\%$ and the statistical term to be $p_1=0.0579\pm0.0006$ MeV$^{1/2}$.

\begin{figure}[!ht]
\begin{center}
\includegraphics [width = 0.45 \textwidth] {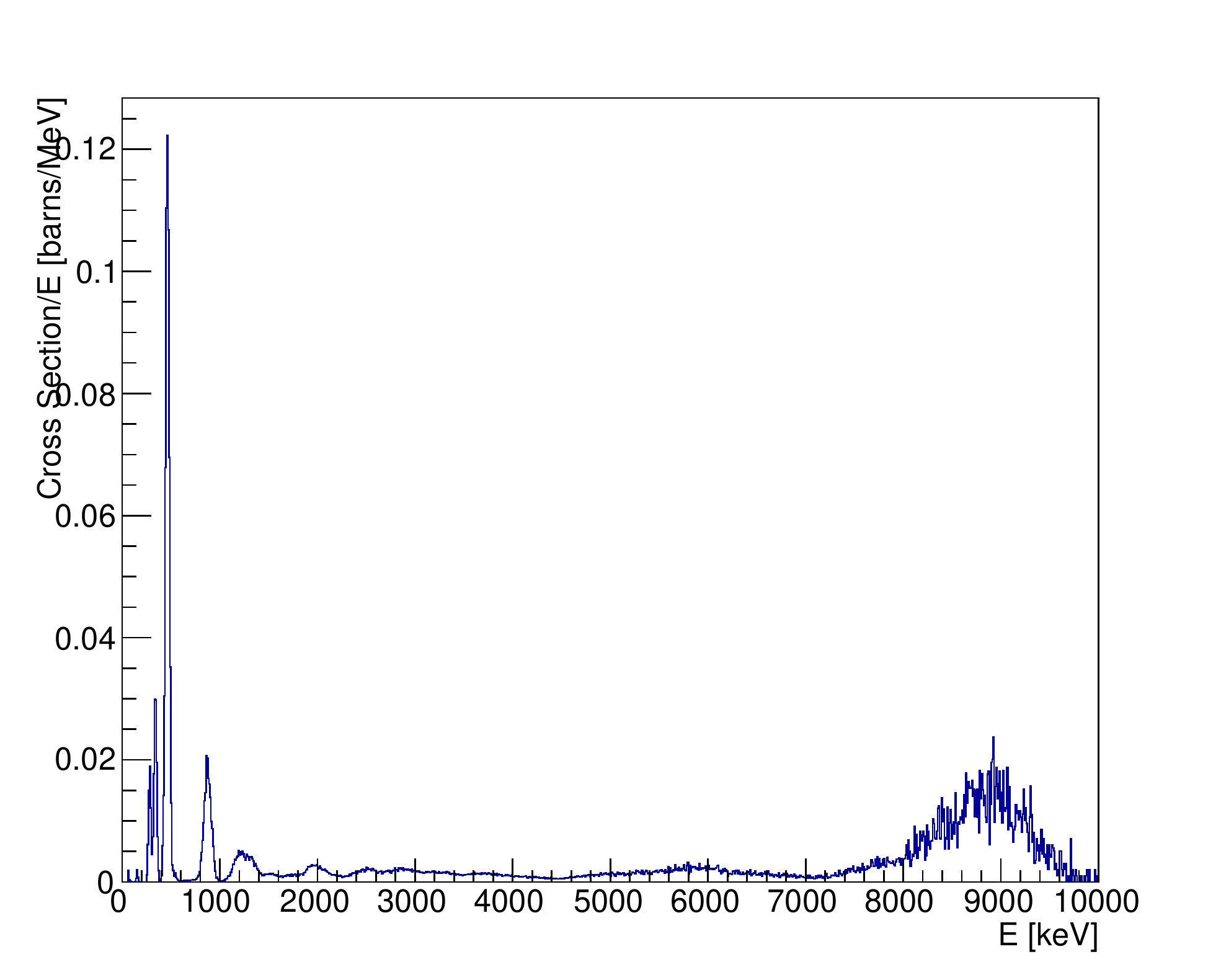}
\caption{ Expected prompt photon spectrum from neutron irradiation of the natural admixtrure of Ni obtained by convoluting the nuclear activation lines as tabulated in Ref.~\protect~\cite{PGNAA} and convolving them with the expected resolution. }
\label{fig:nichel}
\end{center}
\end{figure}

In order to have an additional line for calibration, following Refs~\cite{nichel} we have also inserted a 2mm thick rod of nickel between the AmBe source and the detector. The prompt gamma neutron activation of Ni in fact generates a set of high energy lines that can be used for calibration. In order to understand what to expect in the detector,  from the knowledge of the cross-sections of all the neutron activated prompt photon lines~\cite{PGNAA}, by taking into account the experimental resolution and the natural abundances, we have simulated the expected $\gamma$ spectrum resulting from the neutron irradiation of nickel (see Fig.~\ref{fig:nichel}). There is a dominant structure at high energy  which is due to the superposition of several lines. A Gaussian fit to it returns a mean value of 8795~keV with a 5.2\% width. Since the width is larger than the input resolution, it is dominated by the presence of several lines close-by.

\begin{figure}[!ht]
\begin{center}
\includegraphics [width = 0.45 \textwidth] {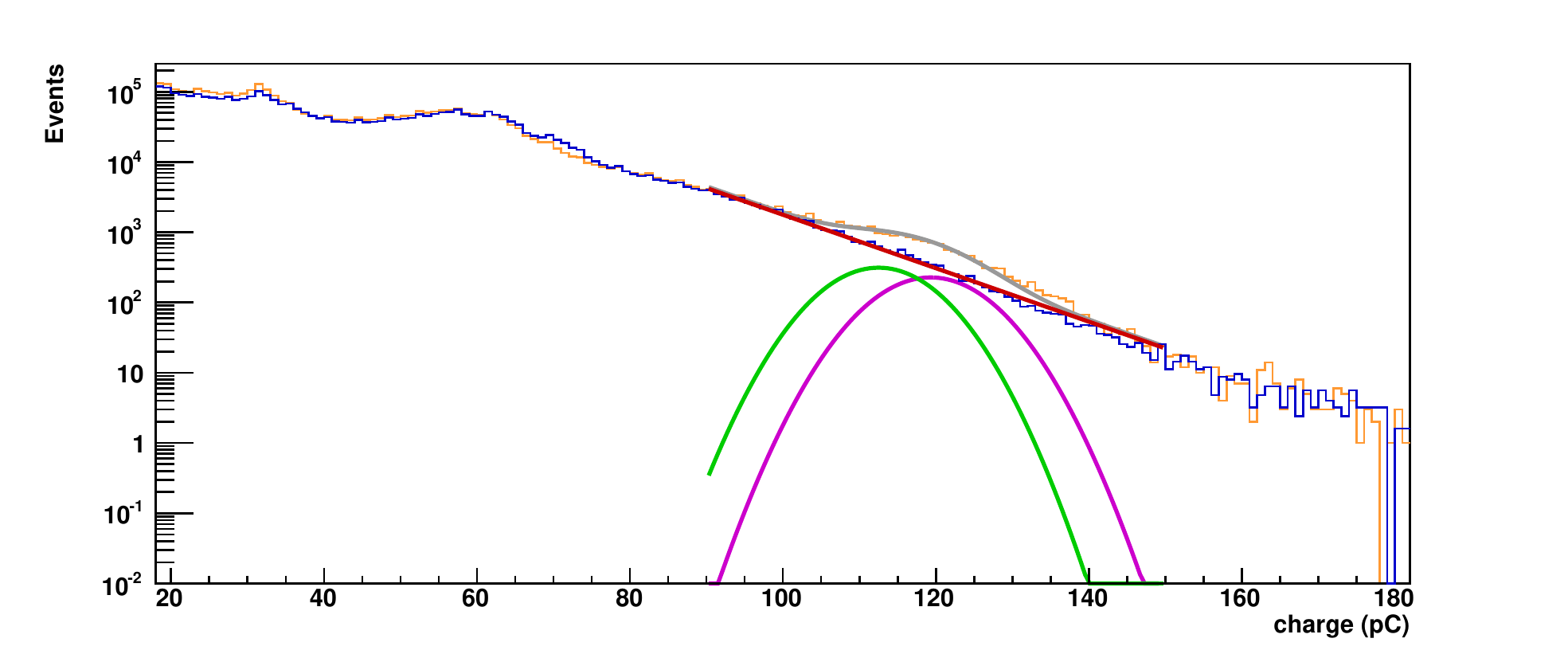}
\caption{Observed spectrum with an AmBe source when a nickel rod is interposed (full orange line) and when it is not (dashed blue line). Fits to both spectra as described in the text are superimposed. In the case of the spectrum with nickel, the Gaussian signals from the structure expected at 8.8~MeV (dotted purple line) and its escape (dashed-dotted green line) are shown. }
\label{fig:nifit}
\end{center}
\end{figure}

Fig.~\ref{fig:nifit} shows the spectra of the recorded charge in two different runs, one when the nickel rod was superimposed (named "nickel" spectrum in the following) and another one taken in the same identical conditions but without the rod  (the "no-nickel" spectrum). The "no-nickel" spectrum is fitted with an exponential curve representing the background. The "nickel" spectrum is fitted with the sum of an exponential with the same slope as the background one and the sum of two Gaussians representing the Ni structure and its escape respectively. The Gaussian widths are fixed to 5.2\% of the mean values, the expected widths as discussed above. The relative offset of the two Gaussians is fixed by the assumption that the escape line is 511~keV below the other one.

Considering the  lines fitted in Figs.~\ref{fig:LYSO2} and~\ref{fig:nifit} and the lines from the $^{22}$Na and $^{60}$Co sources, the calibration plot of Fig.~\ref{fig:LYSO2} is obtained.  A linear dependence of the energy on the  $ADC$ counts is verified on the whole range. The same figure also compares this calibration with the one corresponding to Ref.~\cite{catania} and no significant change is observed.

%% file: Ricalib_simulation.tex
\section{Simulation}
\label{sec:simulation}
Measurements are compared with the 2013.1.0 version of the Monte Carlo code FLUKA~\cite{fluka}. The predefined default 'PRECISIO' is used for the simulation.
With this default, transport options are selected to enable electromagnetic showers, Rayleigh scattering and inelastic form factor corrections for Compton scatterings with Compton profiles activated, 
full analog absorption for low-energy neutrons and restricted ionization fluctuations. A detailed treatment of the photoelectric edge and fluorescence photons is also activated. Thermal neutrons are transported down to $10^{-5}$ eV; other particles are transported down to 100 keV. Delta-ray production threshold is set to 100 keV. Tabulation ratio for hadron and muon dp/dx is set to 1.04; fractional kinetic energy loss per step is set to 0.05. 

The beam is simulated as a cylindrical mono-directional source with a $0.75$ cm lateral radius, located $25\,cm$ away from the target's nearest face. The geometry has been simplified to reproduce the basic elements, target and $LYSO$ detector, which have been described in detail. In order to reduce the contribution of neutrons to the final spectra, a $5$ ns time cut card has been applied, filtering the particles by their time of arrival from the beam exit to the detector. Such value is able to reduce drastically the neutron background. The absence of time structure in the simulation makes it unnecessary to locate a minimum time of flight, as there is no chance of contamination from previous pulses. 
%The Start Counter is not included, so the influence of fast particles coming from this element is not present in the simulation.

%The composition of the LYSO crystal is $Lu_{1.8}Y_{0.2}SiO_5$.

We simulate $6\cdot 10^8$ impinging carbon ions. To increase the statistics of the results, the detector, made of four
LYSO crystals, is virtually replicated in a ring centered in the target
and with axis parallel to the beam, with no possible cross detection
between detectors. At the distance from the target to detector (74cm) the distribution of
particles is practically isotropic, so the energy deposition in the
replicas is summed up to obtain the gamma energy spectra. The replication is taken into account when normalizing to data statistics.
%That way, the convergence of the results is obtained faster, with less histories, remarking the data is later renormalized by the number of replicas utilized.  
Counts in the scintillator are recorded with a DETECT card, which scores energy deposition on an event by event basis, using a 0.2 MeV binning,
to match the experimental data. The scored counts are then folded with the intrinsic resolution of the $LYSO$ crystal as determined in Sec.~\ref{newcalib}.

%% file: Ricalib_results.tex
\section{results}
\label{sec:results}
\begin{figure}[!b]
\centering
\includegraphics[width=3.in]{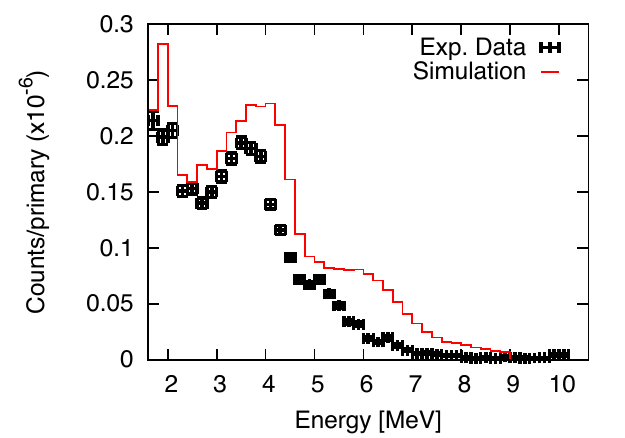}
\caption{Comparison of the calibrated spectrum in data (squares) with the FLUKA simulation(histogram). }
\label{fig:results}
\end{figure}

Fig.~\ref{fig:results} shows the comparison between the prompt photon energy spectrum after calibration and the FLUKA simulation. As it is clear from Fig.~\ref{fig:LYSO2}, the new calibration procedure has confirmed the original calibration and therefore the data spectrum is fully consistent with the published one (Fig.~\ref{fig:catania}). 

\begin{figure}[tbh]
\centering
\includegraphics[width=3.in]{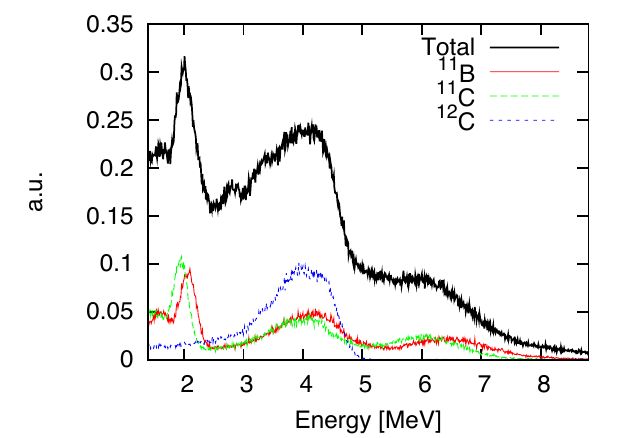}
\includegraphics[width=3.in]{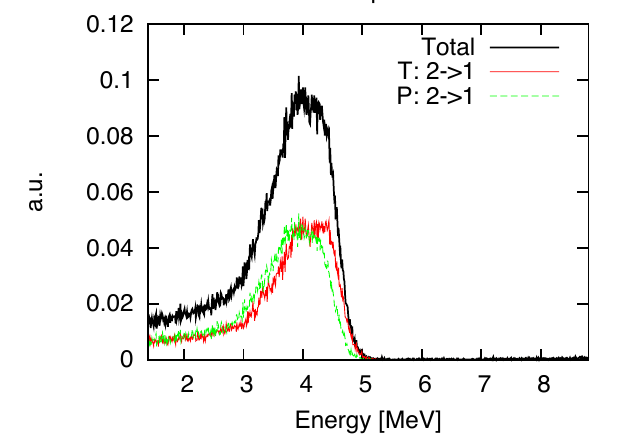}
\caption{MC Study with FLUKA. Top: breakdown of the gamma spectrum among different excited nuclei. Bottom: comparison of the target and projectile contributions in the case of  $^{12}C^*$ de-excitations.}
\label{fig:tot1}
\end{figure}

As far as the simulation is concerned, in our previous publication GEANT showed a discrepancy both in the normalization (a 130\%  excess in simulation) and in the spectrum, in particular in the relative content of the 4.4 MeV line with respect to the rest of the spectrum.
The FLUKA simulation reduces the excess in normalization to $\sim40\%$, and matches the data well as far as the fraction of 4.4MeV photons is concerned. In this respect we have studied the composition of the spectrum.
The main structure in Fig.~\ref{fig:results} is around 4 MeV and it is due to the $^{12}C^*$, $^{11}C^*$ and $^{11}B^*$ lines, mixed as a consequence of the
resolution of the detector. In Fig.~\ref{fig:tot1} the gamma counts from each nuclei are shown. 
Although the  4.4 MeV $^{12}C^*$ line constitutes around the 50\% of the total height of the peak, the two remaining nuclei contribute significantly.
The overall rate of prompt photons with energies above 2 MeV for each isotope is shown in Tab.~\ref{tab:rates}.

%\begin{table}[!bth]
% \begin{center}
%  \begin{tabular}{ccccccc}
%    \hline
%    \hline
%    Energy range & Target $^{11}B$ & Target $^{11}C$ & Target $^{12}C$ & Projectile $^{11}B$ & Projectile $^{11}C$ & Projectile $^{12}C$ \\
%    \hline
%     $[3:5]$      & 7.632    & 6.052  & 15.581  & 9.706  & 9.289  & 14.979\\
%     $[4:5]$      & 9.536    & 6.903  & 16.976  & 10.878 & 9.810  & 13.167\\
%     $[4.2:4.5]$  & 9.884    & 7.973  & 24.185  & 10.904 & 6.435  & 18.947\\
%     $[4.35:4.45]$& 12.624   & 5.338  & 34.639  & 8.139  & 3.442  & 17.978\\
%    \hline
%  \end{tabular}
% \end{center}
% \caption{Fraction of gammas (in \%) selected by their original nuclei. The values are normalized to the total number of counts in the selected range of energies.}
%\end{table}

\begin{table}[!bth]
 \begin{center}
  \begin{tabular}{cc}
    \hline
    \hline
 Target $^{11}B$ & 3.9\\
 Target $^{11}C$ & 3.0\\
 Target $^{12}C$ & 7.5\\
 Projectile $^{11}B$ & 4.8\\
  Projectile $^{11}C$ & 4.5 \\
  Projectile $^{12}C$ & 7.0\\
    \hline
    \hline
  \end{tabular}
 \end{center}
 \caption{\label{tab:rates} Contributions from different excited nuclei to the gamma spectrum. The fraction of gammas with measured energy between 3 and 5 MeV (the so called "$^{12}$C region") with respect to the total number of counts between 2 and 10 MeV.}
\end{table}

It is worth noticing that such gammas can originate both from the  PMMA nuclei de-exciting almost at rest due to peripheral collisions with the beam ions ("target"), or from
projectile  fragments, meaning the de-excitations of the excited beam ions and the subsequent fragments. The latter have a high kinetic energy, which causes Doppler broadening of the  lines. In the bottom of Fig.~\ref{fig:tot1}
the two contributions are compared  in the case of $^{12}C^*$.

As far as there comparison with data is concerned, Tab.~\ref{tab:rates} shows that $^{12}$C$^*$ nuclei  releasing energy in the [3,5] MeV window account for  14.5\% of the prompt photons, number that needs to be compared with the ($13.9\pm0.6$)\% estimated in data~\cite{catania}.
 
There persists instead a 10\% offset in the energy scale, that, given the accuracy of the off-line calibration, is at this point to be attributed to saturation effects that occur at high rate. Such effect is not relevant for the dosimetric application that is the goal of this study, but needs to be accounted for when tuning the simulation to the data.